# Improvement of both performance and stability of photovoltaic devices by in situ formation of a sulfur-based 2D perovskite

*Milon Kundar [a,b], Sahil Bhandari [a,b], Sein Chung [c], Kilwon Cho [c], Satinder K. Sharma [d], Ranbir Singh [d*], and Suman Kalyan Pal [a,b*]*

[a]*School of Physical Sciences, India Institute of Technology Mandi, Kamand, Mandi-175005, Himachal Pradesh, India*

[b]*Advanced Materials Research Centre, India Institute of Technology Mandi, Kamand, Mandi-175005, Himachal Pradesh, India*

[c]*Department of Chemical Engineering, Pohang University of Science and Technology, Pohang - 37673, South Korea*

[d]*School of Computing and Electrical Engineering (SCEE), Indian Institute of Technology Mandi, Kamand, Mandi - 175005, Himachal Pradesh, India*

AUTHOR INFORMATION

**Corresponding Author**

*E-mail: ranbir.iitk@gmail.com, suman@iitmandi.ac.in; Phone: +91 1905 267040

## Abstract

Perovskite solar cells (PSCs) with superior performance have been recognized as a potential candidate in photovoltaic technologies. However, the defects in active perovskite layer induce non-radiative recombination which restricts the performance and stability of the PSCs. The construction of thiophene-based 2D structure is one of the significant approaches for surface passivation of hybrid PSCs that may combine the benefits of the stability of 2D perovskite with the high performance of 3D perovskite. Here, a sulfur-rich spacer cation 2-thiopheneethylamine iodide (TEAI) is synthesized as a passivation agent for the construction of three-dimensional/two-dimensional (3D/2D) perovskite bilayer structure. TEAI-treated

PSCs possess a much higher efficiency (20.06%) compared to the 3D perovskite (MAFAPbI$_3$) devices (17.42%). Time-resolved photoluminescence (TRPL) and femtosecond transient absorption (TA) spectroscopy are employed to investigate the effect of surface passivation on the charge carrier dynamics of the 3D perovskite. Additionally, the stability test of TEAI-treated perovskite devices reveals significant improvement in humid (RH ~ 56%) and thermal stability as the sulfur-based 2D (TEA)$_2$PbI$_4$ material self-assembles on the 3D surface making the perovskite surface hydrophobic. Our findings provide a reliable approach to improve device stability and performance successively, paving the way for industrialization of PSCs.

## Table of Content

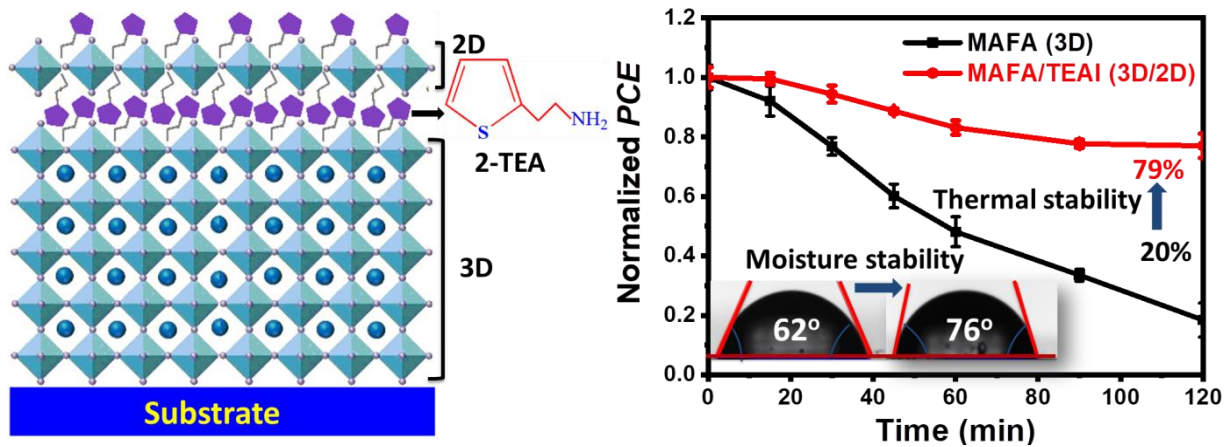

Thiophene-based 2D perovskite is utilized for surface passivation of hybrid perovskite solar cells and acheived improved stability against excessive heat and moisture. Moreover, the surface passivated solar cell exhibits much higher efficiency compared to the 3D perovskite devices.

# 1. Introduction

Organic-inorganic perovskites have appeared as promising semiconducting materials in prospective photovoltaic technologies with low-cost and unique properties.[1, 2] The power conversion efficiency (PCE) of three-dimensional (3D) perovskite solar cells (PSCs) has swiftly improved from 3.8% to 25.7%[3] in the early few years, which is close to silicon-based solar cells. To further improve the efficiency of PSCs several research approaches including composition engineering,[4] perovskite surface passivation,[5-7] crystal growth manipulation,[8, 9] interface engineering,[10, 11] and band alignment[12] are extensively investigated. However, poor device stability under humid, bright, and hot environment conditions is a major obstacle for practical implementation.[13-15]

Recently, two-dimensional (2D) Ruddlesden–Popper (RP) perovskites have stirred research attention for solar cell applications due to their superior environmental stability and structural diversity.[16-18] The structural variation in 2D perovskite is achieved by integrating different cations into the 2D frame.[19] The 2D RP perovskite has a chemical formula $L_2A_{n-1}M_nX_{3n+1}$, where L is the spacer cation such as n-butylammonium ($BA^+$), 2-thiophenemethylammonium ($ThMA^+$), and phenylethylammonium ($PEA^+$) cation, A is the monovalent cation (e.g., formmamidinum ($FA^+$)), M represents divalent metal cation (e.g., $Pb^{2+}$, $Bi^{3+}$), X denotes halide anion, while n indicates the number of confined inorganic $[MX6]^{4-}$ octahedral separated by the spacer. In 2D RP perovskite, the hydrophobic nature of bulky organic spacers protects the inorganic perovskite layer from moisture corrosion. Moreover, the structure steadiness is preserved by van der Waals forces in between the large organic spacers, contributing to the excellent environmental stability.[1] These spacer cations also passivate the defects and improve the ion migration activation energy in 2D perovskites, leading to the enhancement of thermal stability and photostability.[20]

Previous reports suggest that solution-processed perovskite layers are polycrystalline and possess significant structural disorders like grain boundary defects and crystallographic defects.[21-23] Interface defects are another factor that influences the performance of PSCs.[24] One of the most effective strategies for reducing defects is to introduce a 2D passivation layer on a 3D perovskite film. The 2D material could efficiently suppress the film defects, consequently enhance the stability of the PSC by prohibiting moisture penetration into the working layer.[25, 26] This strategy also improves cell efficiency owing to the minimized charge recombination and surface trap states by the appropriate band alignment of 3D/2D perovskite structure.[20, 27] Researchers have attempted surface passivation with low-dimensional perovskite for minimizing surface defects of PSCs, which efficiently enhances device performance. They have modified the perovskite surface by depositing 2D aromatic or aliphatic cations, such as (ThMA$^+$),[16] (PEA$^+$),[20] (BA$^+$),[28] and so on to prepare the 2D perovskite layer.[29, 30] Chen et al. reported an organic halide molecule (i.e., PEAI) for surface defect passivation of triple cation-based perovskite films. The PEAI itself formed a 2D (PEA)$_2$PbI$_4$ perovskite that showed 18.51% efficiency in the solar cell.[20] Zhijun and co-workers synthesized a novel salt 2-thiopheneethylamine thiocyanate (TEASCN) to modify the surface of Sn–Pb based perovskite, and the resulting 3D/2D device exhibited an efficiency of 21.26% owing to remarkably reduced film defects.[31] It has been reported that the thiophene-based quasi-2D RP perovskites show better photovoltaic performances than the phenylethylammonium halide-based perovskite, because of the increased charge mobility provided by the higher polarity of sulfur.[32, 33] Thiophene-based organic amines usually have π−π conjugate structure of aromatic amines. The generation of deep Pb-S interactions and electron-rich π functional group in thiophene, can reduce Pb$^{2+}$ ion defects and inhibit the Pb interstitial formation.[34, 35] Moreover, 2D perovskites with aromatic thiophene ammonium halide cations have stronger hydrophobicity than their aliphatic chain. As a result, the deposition of a thiophene-based 2D perovskite protected the 3D perovskite from deterioration upon aging or heat.[36, 37]

Herein, we synthesize an organic halide salt, 2-thiopheneethylammonium iodide (TEAI), as a basic ingredient for 2D perovskites. MAFAPbI$_3$/(TEA)$_2$PbI$_4$ (3D/2D) bi-layer PSCs were fabricated by introducing TEAI into a 3D perovskite, MAFAPbI$_3$. Our experimental results reveal that the surface passivation with TEAI appreciably improves the optoelectronic properties of the 3D/2D hybrid perovskites, leading to outstanding stability as well as high efficiency of PSCs.

## 2. Experimental section

### 2.1. Materials

2-Thiopheneethylamine (TEA, 98%), lead (II) iodide (PbI$_2$, 99.999%, metals basis), isopropyl alcohol, formamidinium iodide (FAI), methyl ammonium iodide (MAI, 99.9%), dimethylformamide (DMF), dimethyl sulfoxide (DMSO), hydroiodic acid (HI), tin (IV) oxide (SnO$_2$), titanium diisopropoxide bis(acetylacetonate), toluene, anhydrous ethanol, chlorobenzene, acetonitrile, bis(trifluoromethane)sulfonamide lithium salt (Li-TFSI), ethanol, 4-tert-butylpyridine (TBP), tris(2-(1H-pyrazol-1-yl)-4-tert-butylpyridine)cobalt(III) tris(bis(trifluoromethyl sulfonyl)imide) (FK209), 1-butanol, and 2,2′,7,7′-tetrakis-(N,N-di-p-methoxyphenylamine)-9,9′-spirobifluorene (Spiro-OMeTAD) were purchased from Sigma Aldrich. All the materials were utilized without further purification.

### 2.2. Synthesis of TEAI Powder

2-thiopheneethylamine iodide (TEAI) salts were synthesized by reacting the 2-thiopheneethylamine (TEA) with HI (**Figure 1a**). To begin, equimolar TEA and HI were dissolved in ethanol and stirred for at least 2 hours (**Figure 1b**). The solution was then rotary evaporated at 65 °C. After that, TEAI salts were redissolved in ethanol solution and washed numerous times with diethyl ether. Lastly, the white crystals were heated to 65 °C for 6 hours.

## 2.3. Synthesis of $(TEA)_2PbI_4$

A 2D perovskite $(TEA)_2PbI_4$ was prepared following a previous report.[38] We dissolved 0.2 mmol of TEAI and 0.1 mmol of $PbI_2$ with a stoichiometric ratio in 100 μL of DMF to obtain a precursor solution. 50 μL of the perovskite precursor solution was then quickly added to 10 mL of toluene and vigorously stirred. After centrifugation at 5000 rpm for 30 s, 2D perovskite, $(TEA)_2PbI_4$ was obtained.

## 2.4. Device Fabrication

Solar cell structures were fabricated with stack glass /indium-doped tin oxide (ITO)/tin (IV) oxide $(SnO_2)$/MAFAPbI$_3$/(TEA)$_2$PbI$_4$/spiro-MeOTAD/gold (Au) architecture under Ar contained glovebox. First, detergent water, deionized water, acetone, and isopropyl alcohol were used for cleaning ITO substrates in an ultrasonicator consecutively for 20 min. The substrates were then flushed with $N_2$, placed in an oven, and heated for 2 hours at 70 °C before UV-ozone treatment for 20 min. After that, the substrates were spin-casted for 20 s at 2000 rpm with a 15% tin (IV) oxide $(SnO_2)$ colloidal dispersion in $H_2O$ (Alfa-Aesar) and consequently heated at 150 °C for 30 min to generate a compact $SnO_2$ electron transport layer (ETL). For preparing the pristine MAFAPbI$_3$ perovskite film, a precursor solution was prepared by mixing MAI:FAI:PbI$_2$ (1.35 M: 0.15 M:1.5 M) in 4:1 anhydrous DMF:DMSO (1 ml) solvent and stirred whole night at 60 °C. Next, MAFAPbI$_3$ precursor solution was spin-casted via antisolvent technique on the top of ITO/SnO$_2$ layer for 10 s at 1000 rpm, and thus formed perovskite films were annealed at 70 °C for 2 minutes. As an anti-solvent, 0.4 mL toluene was dripped over the film during the last 10 s of coating and then annealed the perovskite films at 70 °C for 2 min. In the TEAI treatment process, TEAI powder was diluted in toluene with four distinct concentrations (0.5, 1, 2, and 3 mg TEAI per 1 mL toluene) and 50 μL of every TEAI solution was used to spin coat onto the MAFAPbI$_3$ perovskite surface at spin rate of 4000 rpm for 30 s and dried for 10 min at 125 °C. To fabricate hole transporting layer (HTL), we doped spiro-OMeTAD (73 mg mL$^{-1}$ in chlorobenzene) with FK209 (18 μL; mother solution: acetonitrile, 300 mg/ml), Li-TFSI (28 μL; mother solution: acetonitrile, 530 mg/ml), and TBP (28 μL) and deposited it on the perovskite film via spin coating for 20 s at 3000 rpm. Lastly, 80 nm thick Au was deposited as a top electrode under the vacuum (< 4 × 10$^{-6}$ Torr) using a mask. We use MAFA and MAFA/TEAI for representing MAFAPbI$_3$ and MAFAPbI$_3$/TEAI perovskites in the following sections.

## 2.5. Characterization

UV-visible absorption and photoluminescence (PL) measurements were performed by Shimadzu UV-2450 spectrometer and HORIBA Fluorolog-3, respectively. We excited the samples at 500 nm for PL measurement. Field-emission scanning electron microscopy (FESEM) images were measured using FEI Nova Nano SEM-450. XRD patterns of perovskite films were acquired by Rigaku Smart Lab 9 kW rotating anode diffractometer. Glazing incidence wide-angle scattering (GIWAXS) measurements were conducted at Pohang Light Source (PLS-II), South Korea. The GIWAX photographs were taken at 0.13 incidence angle using 11.57 keV X-rays (λ = 1.0716Å) as well as a MAR345 imaging plate detector. Ultraviolet photoelectron spectroscopy (UPS) measurements were executed using Nexsa base equipment. J-V measurements were carried by a Keithley 2400 source meter under irradiation intensity of 1 sun at AM 1.5G (100 mW cm$^{-2}$) at room temperature. The external quantum efficiency (*EQE*) measurements were conducted employing the EQE instrument (Zolix Instruments, Inc). Time-resolved photoluminescence (TRPL) measurements were performed using HORIBA DeltaFlex-011x system with a LED source (wavelength 454 nm). Femtosecond transient absorption (TA) measuring setup uses a Ti- sapphire regenerative amplifier (spitfire ace, spectra physics) which is seeded by a laser oscillator (Mai Tai SP, spectra physics). The output of the amplifier (wavelength ~ 800 nm, pulse width < 35 fs and energy 4 mJ per pulse) was divided to generate pump and probe pulses. The pump pulse at 532 nm and probe pulse at 750 nm were obtained from the nonlinear optical parameter amplifier (TOPAS). A small fraction of 800 nm was used to generate probe light in the region 440-770 nm, with the help of sapphire crystal. TA spectra were recorded by dispersing the probe beam with a grating photograph (Acton spectra pro SP 2358) followed by a CCD array. Group velocity dispersion of probe beam was compensated by using a chirp correction program (Pascher instrument). Two photodiodes of variable gain were used two detect the transient absorption kinetics. A DSA100 analyzer was used to measure the contact angle of a water droplet.

## 3. Results and Discussion

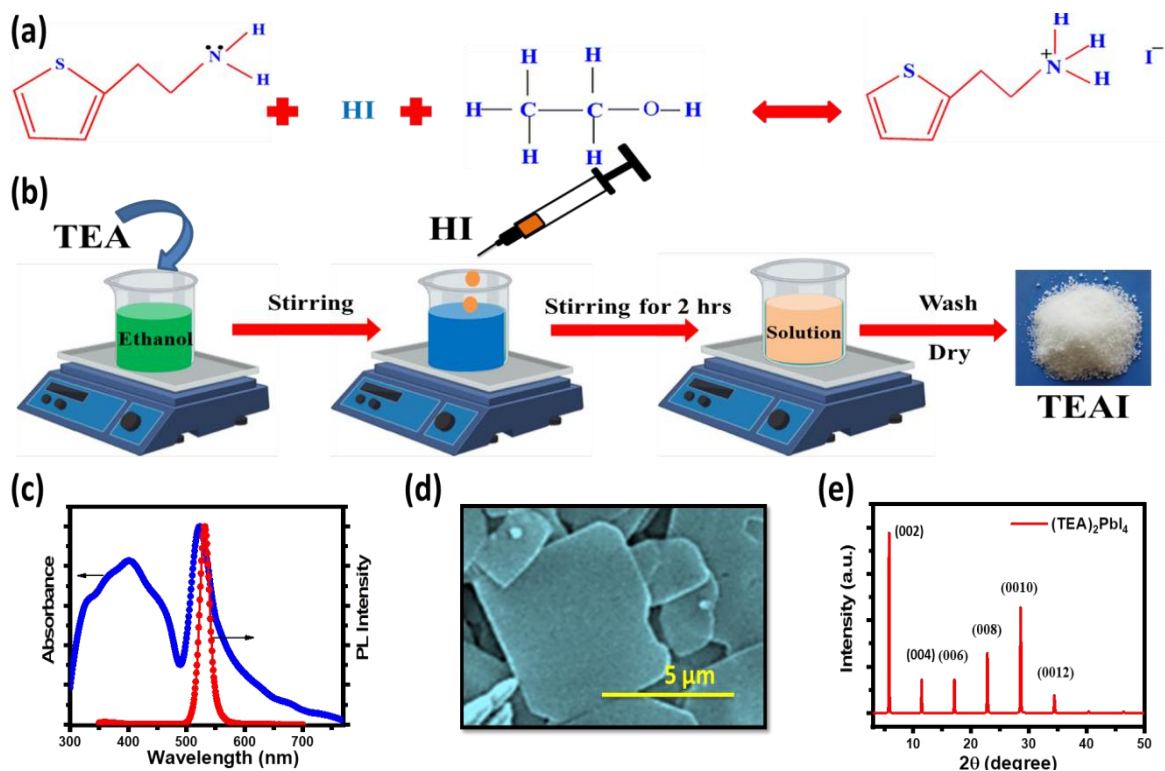

**Figure 1.** (a) Synthetic route of TEAI, (b) schematic diagram of the synthesis process of TEAI. (c) UV-vis absorption and photoluminescence (PL) spectra, (d) scanning electron microscopy (SEM) image, and (e) X-ray diffraction (XRD) pattern of 2D $(TEA)_2PbI_4$ perovskite films.

To obtain aromatic ring-based monovalent TEA$^+$ cation, we synthesized sulfur-based molecule 2-thiopheneethylamine iodide (TEAI) from TEA and HI (see experimental section). We conducted X-ray diffraction (XRD) of TEAI powder and film on the glass substrate (**Figure S1**). The presence of a low angle peak (2θ ≈ 8.7°) in both XRD patterns implies 2D crystal structure. Compared to the TEAI powder sample, the film exhibits a very low-intensity diffraction peak. Then, we synthesized 2D perovskite, (TEA)$_2$PbI$_4$ using TEAI (see experimental section). UV-visible absorption and PL spectra were measured to explore the optical properties of (TEA)$_2$PbI$_4$ thin films. **Figure 1c** shows two main absorption peaks located at 401 nm (3.09 eV) and 521 nm (2.38 eV). The broad peak at 401 nm could be ascribed to transitions to higher energy levels, whereas the narrow absorption peak at 521 nm is attributed to the absorption due to the intrinsic exciton in the generated quantum well structure.[39] The PL peak of (TEA)$_2$PbI$_4$ is centered at 530 nm (2.33 eV) with a full-width at half-maximum (FWHM) of 21 nm. The narrow bandwidth of the PL spectrum reveals that the emission arises from the excitonic recombination.[40] We employed scanning electron microscopy (SEM) to investigate the surface morphologies of (TEA)$_2$PbI$_4$ crystals. The obtained 2D crystals are typically truncated rectangular shapes having lateral dimensions up to several micrometers (**Figure 1d**). XRD of the thin film sample was studied to identify the structures of (TEA)$_2$PbI$_4$ crystal and presented in **Figure 1e**. The crystal possesses a 2D perovskite structure as evidenced by the low-angle peak for the plane (002) at 2θ ≈ 5.8°. The 2D (TEA)$_2$PbI$_4$ shows diffraction peaks with spacings at 5.8°, 11.45°, 17.12°, 22.83°, 28.60°, and 34.44°, which could be assigned to the (00ℓ) lattice planes (ℓ denotes an even number). The XRD result is found to be similar to that reported for 2D layered perovskite.[41, 42] Moreover, the as-prepared TEA$_2$PbI$_4$ crystals exhibit narrow and strong diffraction peaks, suggesting that the material has high crystallinity and large grain size.

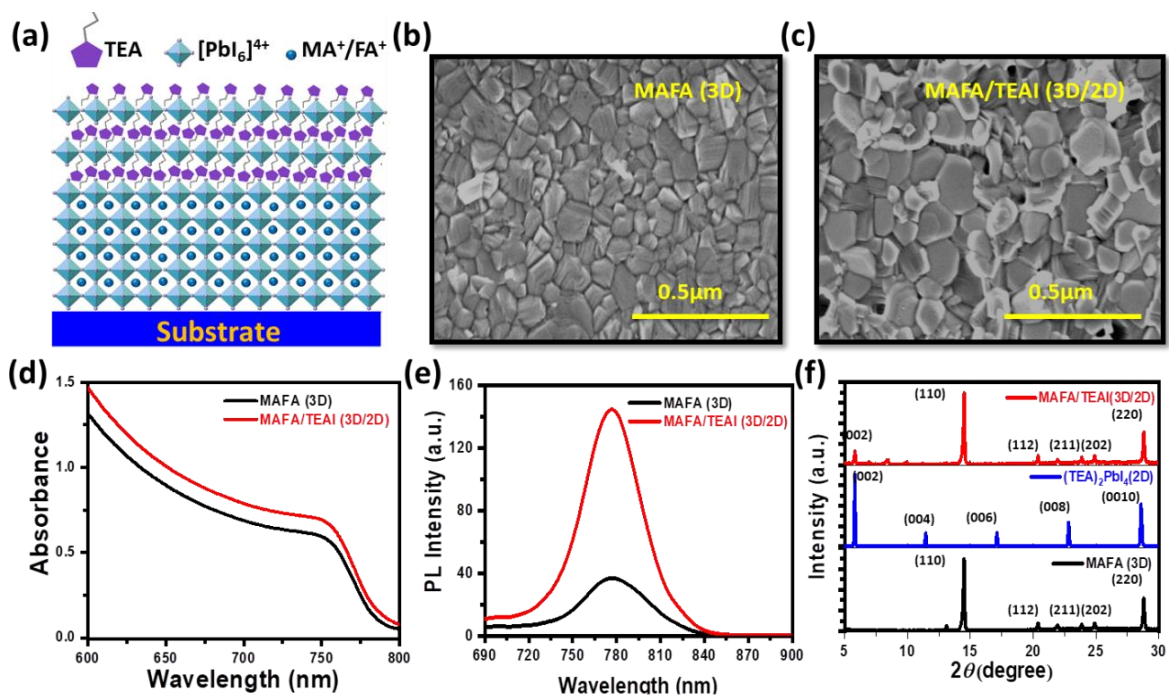

**Figure 2.** Surface passivation of MAFA (3D) layer by TEAI. (a) Schematic diagram of MAFA/TEAI (3D/2D) perovskite structural growth due to TEAI surface treatment. Top-view of SEM images of (b) MAFA (3D) and (c) MAFA/TEAI (3D/2D) films. (d) UV−vis absorption spectra, and (e) PL spectra of MAFA (3D) and MAFA/TEAI (3D/2D) perovskite films. (f) XRD patterns of MAFA (3D), $(TEA)_2PbI_4$ (2D), and MAFA/TEAI (3D/2D) perovskite films.

In this study, we employed 3D perovskite $MAFAPbI_3$ as the light-absorbing layer, prepared by spin-coating method. Then a facile technique was introduced, where TEAI powder dissolved in toluene was spin-casted onto the MAFA (3D) perovskite surface. To test the effect of surface passivation, four different concentrations (0.5, 1, 2, 3 mg TEAI per 1mL toluene) of TEAI were used to spin coat onto the 3D control films. **Figure 2a** shows the possible surface passivation mechanism of MAFA (3D) layer by TEAI molecule. The top-view SEM images of MAFA (3D) and MAFA/TEAI (3D/2D) films coated on $ITO/SnO_2$ are presented in **Figure 2(b-c)**. As compared to the reference 3D film, the TEAI-treated perovskite film possesses a larger grain size indicating that the TEAI induces more crystallinity in perovskite layer. We optimized the effect of the concentration of TEAI on the surface morphology of the perovskite film. It is found from **Figure S2** that the grain size grows with increasing concentration of TEAI, but cracks and holes on the film surface appear at a concentration of 3 mg/mL. The absorption spectra of control and TEAI passivated films were examined to investigate the semiconductor properties of the films, as displayed in **Figure 2d**. It is interesting to note that the absorption intensity for TEAI-modified film is increased over the entire wavelength region (**Figure S3**), which could be attributed to the reduced light scattering caused by the smooth and large grain size of 3D/2D film.[43] We examined PL spectra of MAFA (3D) and MAFA/TEAI (3D/2D) perovskite films as shown in **Figure 2e**. We observe that the PL intensity of the perovskite film is increased after TEAI treatment due to increased absorption of MAFA/TEAI film. Nonetheless, the considerable reduction of nonradiative recombination in passivated layers cannot be ruled out.[44, 45] The PL peaks of TEAI-coated films are slightly shifted from MAFA (3D) due to the reaction between TEAI molecule and excess $PbI_2$ in MAFA (3D) (**Figure S4**). **Figure 2f** reveals XRD patterns of MAFA (3D), $(TEA)_2PbI_4$ (2D), and MAFA/TEAI (3D/2D) perovskite layers. A new peak form at a smaller angle (approximately 5.8°) that is consistent with 2D $TEA_2PbI_4$ perovskite and no apparent peak is at 8.7°, which corresponds to the TEAI powder or film (**Figure S1**). This result infers the presence of $TEA_2PbI_4$ on the surface of $MAFAPbI_3$ film after passivating with TEAI. Therefore, TEAI treatment of the 3D perovskite film leads to the formation of a 3D/2D heterojunction.

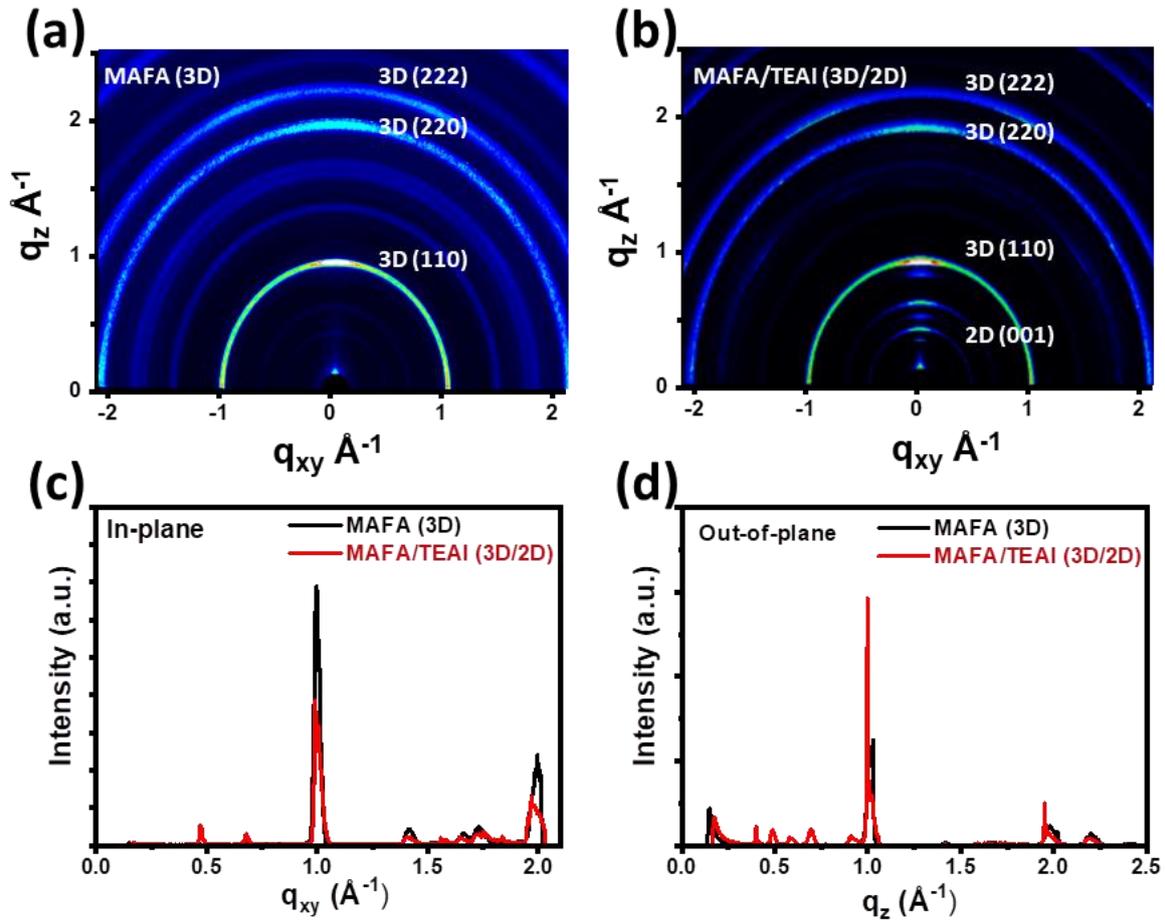

**Figure 3.** (a-b) Grazing-incidence wide-angle X-ray scattering (GIWAXS) images of MAFA (3D) film without and with TEAI treatment. (c) In-plane line cut, and (d) out-of-plane line cut profiles of the control and TEAI-treated films.

The crystallinity difference between MAFA (3D) and MAFA/TEAI (3D/2D) layers was explored by grazing-incidence wide-angle X-ray scattering (GIWAXS). **Figure 3(a-d)** presents the GIWAXS profiles of the MAFA (3D) and interface passivated perovskite films. We observe the isotropic diffraction peak of 3D layer at $q \approx 1.0$ and $1.99$ Å$^{-1}$, corresponding to (110) and (220) crystal planes in both films.[46] We also observe another diffraction pattern at $q_z \approx 0.9$ Å$^{-1}$ associated with the typical peak of PbI$_2$. With the incorporation of TEAI, a small amplitude peak at $q_z \approx 0.4$ Å$^{-1}$ emerged which is attributed to the formation of 2D perovskite on 3D layer. Moreover, several distinct peaks are appeared in 3D/2D layers along out-of-plane at lower angle ($q < 0.75$ Å$^{-1}$) that also reported to the 2D layered material.[47] After treatment with TEAI, the intensity of (110) plane of the perovskite along out-of-plane is significantly increased (**Figure 3d**), indicating improved crystallinity of the films, which can contribute to enhance the PCE and stability of PSCs.[48]

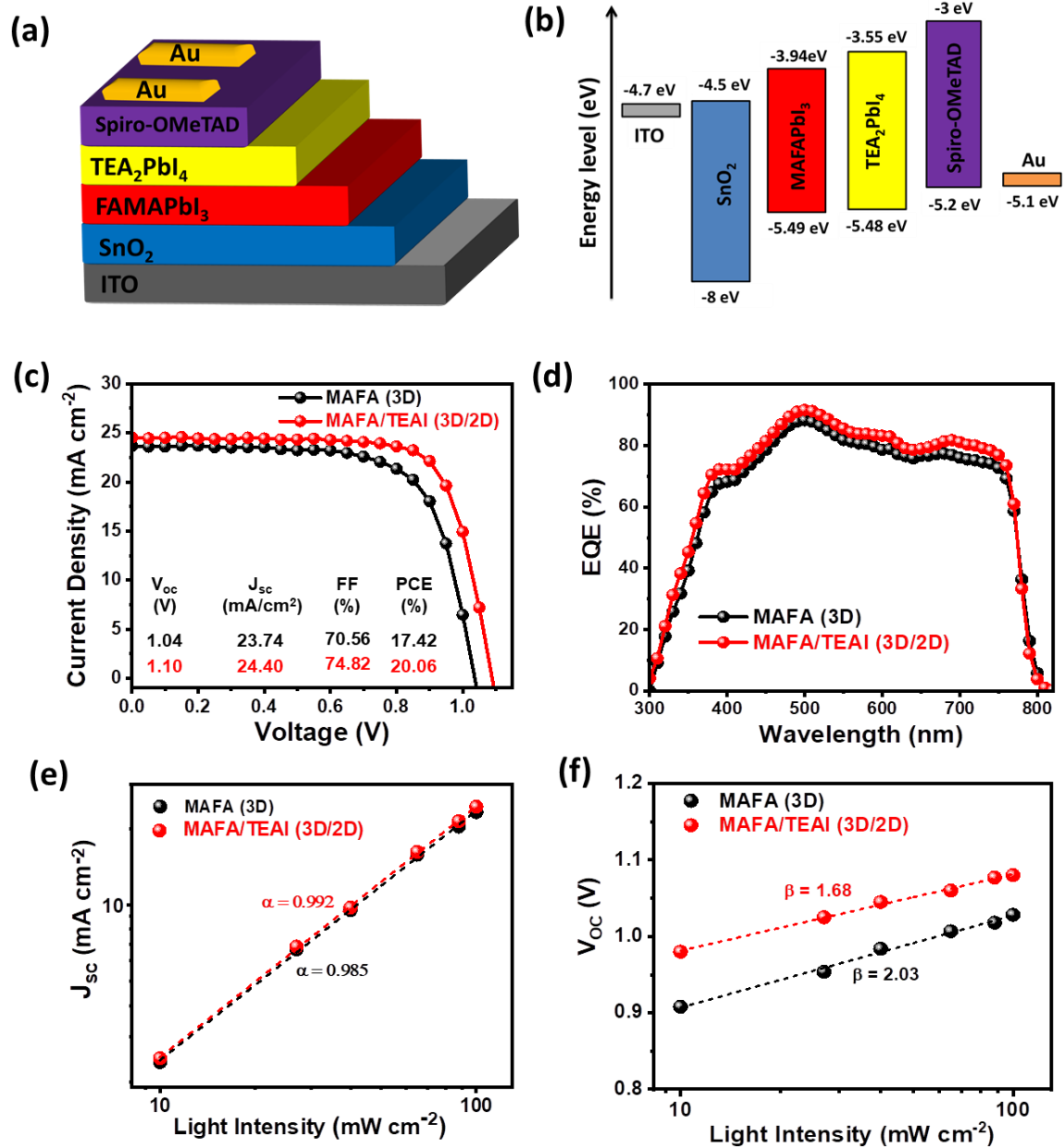

**Figure 4.** (a) Schematic diagram of the device structure, and (b) energy level diagram of every component in the PSC. (c) The typical current density–voltage (J–V) curve of the cells with TEAI (1 mg/mL) and without TEAI treatment under one-sun (100 mWcm$^{-2}$). (d) External

quantum efficiency (EQE) spectra for MAFA (3D), and MAFA/TEAI (3D/2D) PSCs. (e) Plot of $J_{SC}$, and (f) $V_{OC}$ versus light intensity for MAFA (3D), and MAFA/TEAI (3D/2D) PSCs.

Next, we assessed the photovoltaic performance of the PSCs without and with TEAI treatment. **Figure 4(a)** and **(b)** show the device structure and energy band diagram of the components of the PSCs respectively that we adopted in our study. The valence band maximum (VBM) of $TEA_2PbI_4$ perovskite was estimated by ultraviolet photoelectron spectroscopy (UPS). As illustrated in **Figure S5**, the VBM is 5.48 eV and the conduction band minimum (CBM) is 3.55 eV that we calculated from VBM and the optical bandgap (1.93 eV). The energy level diagram suggests that energy levels of the 2D perovskite match well with the pristine MAFA (3D) perovskite for facilitation charge separation. We also compare the energy levels of $TEA_2PbI_4$ and $PEA_2PbI_4$ perovskites (**Figure S6**) and found that $TEA_2PbI_4$ has more suitable energy levels compared to $PEA_2PbI_4$ for charge transport within the solar cell. This implies that 2D $TEA_2PbI_4$ perovskite could be potential better photoactive material for MAFA based 3D/2D solar cells. We measured the performance of PSCs having different concentrations of TEAI solution (**Table S1**) and observed that the maximum efficiency is achieved at a concentration of 1 mg/mL of TEAI. **Figure 4c** presents the current density–voltage (J–V) graphs of the best-performing cells without and with TEAI treatment. The control cell has a maximum PCE of 17.42% with a $J_{SC}$ of 23.74 mAcm$^{-2}$, a $V_{OC}$ of 1.06 V, and a fill factor (FF) of 70.56%. After optimizing the 2D-treated devices, the highest PCE of 20.06% with a $J_{SC}$ of 24.4 mA cm$^{-2}$, a $V_{OC}$ of 1.11 V, and 74.82% FF (**Table 1**) is achieved. The ability of our fabricated solar cells to collect light over a broad wavelength range and generate charge carriers with high efficiency were confirmed by monitoring their external quantum efficiency (EQE) spectra. **Figure 4d** depicts the impact of TEAI on the EQE spectrum of fabricated PSCs. The devices with TEAI treatment show a significant increase in photo-response over the entire wavelength range, and EQE approaches the highest value of 93%. The higher EQE for 3D/2D PSC compared to control PSC could be attributed to improved film quality with decreased defects.

**Table 1.** Photovoltaic parameters of solar cells without and with TEAI treatment

| Perovskite | $V_{OC}$ (V) | $J_{SC}$ (mA/cm$^2$) | FF (%) | PCE (%) | PCE$_{max}$ (%) |
|---|---|---|---|---|---|
| MAFA (3D) | 1.04 ± 0.02 | 23.58 ± 0.16 | 69.93 ± 0.63 | 17.15 ± 0.27 | 17.42 |
| MAFA/TEAI (3D/2D) | 1.10 ± 0.01 | 24.28 ± 0.12 | 74.21 ± 0.61 | 19.82 ± 0.24 | 20.06 |

To further analyze the carrier recombination in control and TEAI-treated devices, dependence of J−V features on light intensity (*I*) was assessed (**Figure S7**). We varied the incident light intensity from 100 mWcm$^{-2}$ (1 Sun) to 10 mWcm$^{-2}$ (0.1 Sun). **Figure 4e** depicts a relationship ($J_{SC} \propto I^\alpha$) between $J_{SC}$ and I on a double-logarithmic scale. The index α is near to 1 when a PSC exhibits weak space charge effects, and most of the charges are collected by electrodes before recombination.[49] The α-value for both 3D and 3D/2D devices is high (0.992 and 0.985 respectively), indicating smooth carrier transport in both devices. Despite the addition of insulating TEAI, no apparent charge barrier at the interface is observed. Furthermore, the dependence of $V_{OC}$ on I for PSCs with and without TEAI is illustrated in **Figure 4f**. The relationship between $V_{OC}$ and $J_{SC}$ can be expressed as follows

$$V_{OC} = \frac{\beta k_B T}{q} ln \frac{J_{SC}}{J_o} + 1 \qquad (1)$$

where β is the ideal factor, $k_B$ is the Boltzmann constant, q is electric charge, T is temperature, and $J_o$ is saturated current density in dark. Assuming $J_{SC} \propto$ light intensity (I) and $J_{SC} >> J_o$, the above formula can be simplified as

$$V_{OC} \propto \frac{\beta k_B T}{q} \ln I \qquad (2)$$

Hence, the plot of $V_{OC}$ with ln*I* provides a slope of $\frac{\beta k_B T}{q}$. It is known that trap-assisted recombination plays a significant role if the slope shifts from $k_B T/q$.[50] The MAFA (3D) device exhibits a slope of 2.03$k_BT/q$, while the TEAI-passivated device shows a much lower slope (1.68$k_BT/q$), suggesting reduced trap-assisted recombination in TEAI-treated solar cells, resulting enhanced device efficiency.

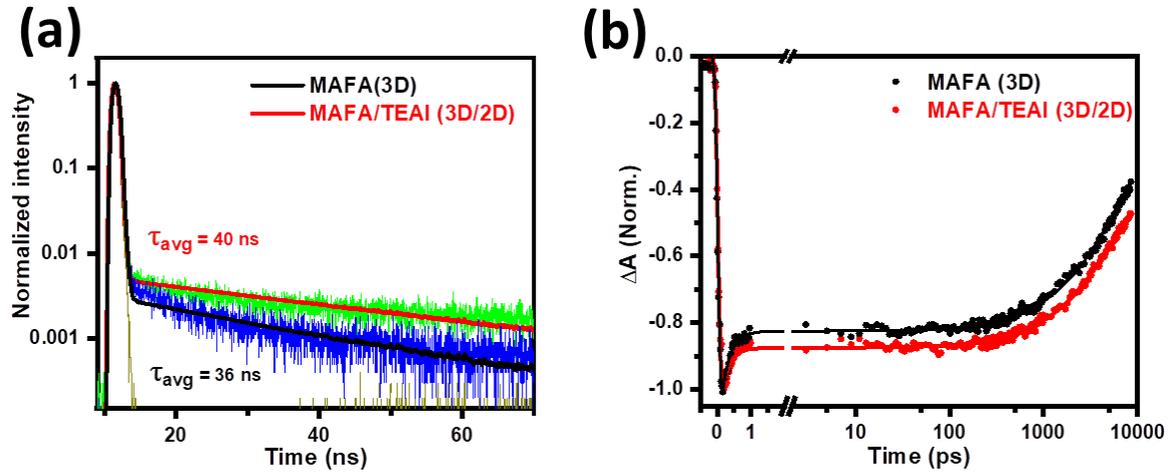

**Figure 5.** (a) Time-resolved photoluminescence (TRPL) of MAFA (3D) and MAFA/TEAI (3D/2D) perovskite films. (b) Comparison of normalized transient absorption (TA) kinetics of MAFA (3D) (red) and MAFA\TEAI (3D\2D) (black) (solid lines are fitting results).

We conducted TRPL measurements to investigate the carrier dynamics in a passivated perovskite layer. The decay curves presented in **Figure 5(a)** were fitted by a bi-exponential function [51]

$$F(t) = \sum_{i=1}^{2} f_i \exp(-t/\tau_i) \qquad (3)$$

where $f_i$ indicates the contribution of the i[th] decay component having lifetime $\tau_i$. PL decay kinetics are fitted well with equation 3 and the fitting results are presented in **Table 2**. According to previous literature, the fast decay component ($\tau_1$) arises from surface trap density, and the slow time component ($\tau_2$) is associated with the recombination that takes place in the bulk perovskite structure.[52] The average PL lifetime ($\tau_{avg}$) can be calculated from the following equation [53]

$$\tau_{avg} = \frac{\sum_i f_i \tau_i}{\sum_i f_i} \qquad (4)$$

**Table 2.** PL lifetimes extracted from fitting PL decay curves with a bi-exponential decay function

| Perovskite | $\tau_1$ (ns) | $f_1$ (%) | $\tau_2$ (ns) | $f_2$ (%) | $\tau_{avg}$ (ns) |
|---|---|---|---|---|---|
| MAFA (3D) | 0.15 | 94 | 26.6 | 6 | 36 |
| MAFA/TEAI (3D/2D) | 0.1 | 87 | 41.2 | 13 | 40 |

The average PL lifetime are calculated to be 36 and 40 ns for MAFA and MAFA/TEAI films, respectively. The prolonged carrier life could be explained by the reduced defect-assisted nonradiative recombination as consequence of high crystallinity and low concentration of defects in MAFA/TEAI.[52] In the interface passivated film, $\tau_1$ is significantly decreased indicating the reduction of trap-assisted recombination. Additionally, the slower recombination caused by appreciably improved $\tau_2$ in 3D/2D film surely contributes to the device performance with higher FF and $J_{SC}$ due to the sufficient charge collection as observed from the J-V graphs. The findings of TRPL measurements confirm that interface modification of MAFA (3D) by 2D (TEA)$_2$PbI$_4$ perovskite has a significant effect on charge carrier lifetime in solar cell devices.

To further examine the effect of TEAI passivation, we investigated the charge carrier dynamics of MAFA (3D) using femtosecond TA spectroscopy. Perovskite samples were photoexcited by a pump pulse at 532 nm and white light continuum (WLC) was used to probe the sample. **Figure S8(a)** shows TA spectra of MAFA (3D) film with a negative absorption band in the wavelength region 730-770 nm. It should be noted that we were unable to capture complete TA band due to week probe light after 770 nm. As MAFA (3D) has ground state absorption in this wavelength region (**Figure 2(d)**), the negative TA band could be assigned to the ground state bleach (GSB) which arises due to the state filling of the band edge states.[54] Next, we measured TA spectra of TEAI passivated 3D perovskite film, which shows a GSB band (**Figure S8(b)**) similar to MAFA (3D). To estimate lifetimes of photogenerated carriers, we measured TA kinetics of MAFA (3D) and MAFA\TEAI (3D\2D) films at GSB band (**Figure 5(b)**). Clearly, TA decay is slower in MAFA\TEAI (3D\2D) suggesting longer charge carrier lifetime in that sample. TA kinetics were fitted with a biexponential function to find average lifetimes (**Table 3**). The fitting results infer that the average lifetime of charge carriers in MAFA perovskite increases after surface passivation with TEAI which is consistent with the TRPL results. Therefore, the enhanced performance of the 3D/2D perovskite based solar cells could be ascribed to the slow relaxation of charge carriers in MAFA\TEAI (3D\2D).

**Table 3.** Results of biexponential fitting of TA kinetics

| Perovskite | $t_1$ (ps) | $f_1$ (%) | $t_2$ (ps) | $f_2$ (%) | $<\tau>^*$ (ns) |
|---|---|---|---|---|---|
| MAFA (3D) | 0.25 | 37 | 5784 | 63 | 3.6 |
| MAFA/TEAI (3D/2D) | 0.23 | 31 | 5850 | 69 | 4 |

$^*<\tau> = f_1 t_1 + f_2 t_2$

We examine the thermal stability of the PSCs without and with TEAI passivation by damp heating test using a hot plate at 80 °C temperature in the ambient atmosphere. We measured the photovoltaic performances of control, and 3D/2D non-encapsulated PSCs in every 15-25 min and the values of different device parameters are presented in **Figure 6(a-d)**. Under continuous heat treatment at 80 °C for 2 h, initial performance parameters of the 3D device are dropped by a great extent, while the TEAI-treated device exhibit much better performance than untreated one. The treated device maintains almost constant PCE (79% of the initial PCE) after an initial drop under similar conditions. 2D material has a relatively higher binding energy than 3D, which requires much more activation energy to stimulate ion migration when heated.[55, 56] As a thermally stable layer, the (TEA)$_2$PbI$_4$ 2D layer could effectively suppress the iodide ion diffusion from MAFA (3D) layer to the HTL and electrode, significantly improving the thermal stability of PSCs. To monitor operational stability, all devices were subjected to one-sun (100 mWcm$^{-2}$) illumination continuously for 500 s. **Figure S9** shows the time-dependent short circuit current density ($J_{SC}$), and open-circuit voltage ($V_{OC}$) of the PSCs without and with TEAI passivation. The $V_{OC}$ of PSCs with TEAI-treatment provides a steady $J_{SC}$ for 500 s under constant light irradiation, whereas it decays constantly for MAFA (3D) device. The $J_{SC}$ of TEAI-treated devices drop down from 24 mA/cm$^2$ to 13 mA/cm$^2$, while the J$_{SC}$ of pristine cell reduces drastically to 7 mA/cm$^2$ within 500 s. This stability improvement in TEAI-passivated PSCs under constant light irradiation can be attributed to the improved film morphology.

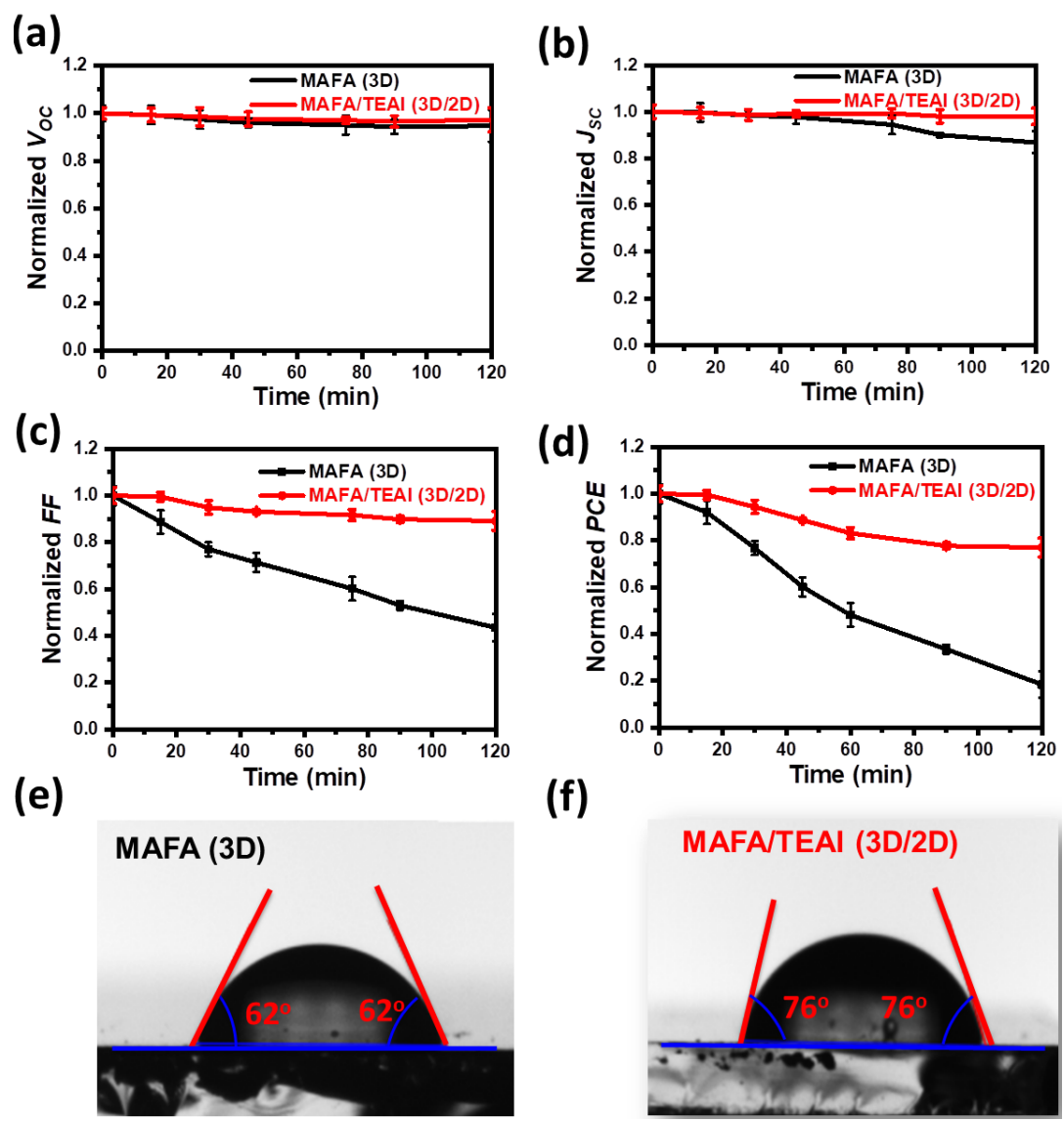

**Figure 6.** Photovoltaic parameters of unsealed PSCs heating on a hot plate at 80 °C for 2 h. Normalized (a) short circuit current density (J$_{SC}$), (b) open-circuit voltage (V$_{OC}$), (c) fill factor (FF), and (d) power conversion efficiency (PCE) of the photovoltaic devices without and with TEAI. Images showing the contact angles of a water droplet on (e) MAFA (3D), and (f) MAFA/TEAI (3D/2D) perovskite films.

Finally, the moisture stability of pristine and TEAI passivated devices was assessed by contact angle measurement of water droplet (**Figure 6(e-f)**). The contact angle for MAFA (3D) film is 62°, whereas it increases to 76° for the interface passivated film which exhibits the highest device performance. This result confirms the wetting of passivated perovskite surface by water is substantially decreased, indicating an enriched hydrophobicity. Therefore, TEAI-treated film could exhibit higher moisture stability than 3D film because of the hydrophobicity of 2D materials. Moreover, the contact angle rises gradually up to 83° when the concentration of TEAI goes up, inferring that the 2D $(TEA)_2PbI_4$ perovskite phase coverage grows.[57] We also tested the water droplet contact angle on PEAI-treated perovskite film for comparison and found to be 67° (**Figure S10**). It is interesting to note that the presence of TEAI on 3D perovskite significantly improves the hydrophobicity of the films rather than PEAI on the same 3D perovskite and thus boosts the stability of PSCs.

## 4. Conclusions

In conclusion, we have demonstrated an effective strategy for passivation of 3D $FAMAPbI_3$ perovskite surface with a 2D $(TEA)_2PbI_4$ layer to acquire highly efficient and stable 3D/2D PSCs. We have found that the TEAI molecule is successfully embedded into $FAMAPbI_3$ to grow a 2D thin layer on the 3D surface. The incorporation of thiophene-based cation on the perovskite film substantially reduces the defects, leading to a longer carrier lifetime and lower nonradiative recombination as confirmed by TRPL and TA measurements. As a consequence, the TEAI-treated device achieved a champion PCE of 20.06% along with a high $V_{OC}$ of 1.11 V, exceeding that (PCE = 17.42%, $V_{OC}$ = 1.06 V) of the control device. Importantly, the enhanced hydrophobicity due to the thiophene-based 2D perovskite layer contributes to the improved stability against excessive heat and moisture (RH ≈ 56% ± 4%). The optimized devices maintained approximately 79% of their earlier PCE after 2 h under continuous heat treatment at 80 °C, while the 3D devices lost majority of their initial performance. It is also observed that $(TEA)_2PbI_4$ perovskite has a more suitable band alignment and higher moisture stability compared to the earlier used (PEA)$_2$PbI$_4$ perovskite. Our findings demonstrate that 2D (TEA)$_2$PbI$_4$ perovskite can be an excellent candidate for developing highly stable and efficient PSCs.

## Acknowledgments

We are thankful to Indian Institute of Technology Mandi for providing the experimental facilities at Advanced Material Research Centre (AMRC) and Centre for Design and Fabrication of Electronic Device (C4DFED). M. K. acknowledges the Ministry of Human Resource Development (MHRD) for his fellowship. R. S. wishes to thank the Science and Engineering Research Board (SERB), New Delhi, for the prestigious Ramanujan Fellowship, 2020 (grant no. RJF/2020/000005). Portions of this research were carried out at the 3C and 9A beam lines of the Pohang Accelerator Laboratory, Republic of Korea.

## References

1. Lai, H.; Lu, D.; Xu, Z.; Zheng, N.; Xie, Z.; Liu, Y., Organic-Salt-Assisted Crystal Growth and Orientation of Quasi-2D Ruddlesden–Popper Perovskites for Solar Cells with Efficiency over 19%. *Adv. Mater.* **2020,** *32* (33), 2001470.

2. Jiang, X.; Chen, S.; Li, Y.; Zhang, L.; Shen, N.; Zhang, G.; Du, J.; Fu, N.; Xu, B., Direct surface passivation of perovskite film by 4-fluorophenethylammonium iodide toward stable and efficient perovskite solar cells. *ACS Appl. Mater. Interfaces* **2021,** *13* (2), 2558-2565.

3. Kim, M.; Jeong, J.; Lu, H.; Lee, T. K.; Eickemeyer, F. T.; Liu, Y.; Choi, I. W.; Choi, S. J.; Jo, Y.; Kim, H.-B., Conformal quantum dot–SnO2 layers as electron transporters for efficient perovskite solar cells. *Science* **2022,** *375* (6578), 302-306.

4. Kang, D. H.; Park, N. G., On the current–voltage hysteresis in perovskite solar cells: dependence on perovskite composition and methods to remove hysteresis. *Adv. Mater.* **2019,** *31* (34), 1805214.

5. Aydin, E.; De Bastiani, M.; De Wolf, S., Defect and contact passivation for perovskite solar cells. *Adv. Mater.* **2019,** *31* (25), 1900428.

6. Jacobsson, T. J.; Correa-Baena, J.-P.; Halvani Anaraki, E.; Philippe, B.; Stranks, S. D.; Bouduban, M. E.; Tress, W.; Schenk, K.; Teuscher, J. l.; Moser, J.-E., Unreacted PbI2 as a double-edged sword for enhancing the performance of perovskite solar cells. *J. Am. Chem. Soc.* **2016,** *138* (32), 10331-10343.

7. Zheng, X.; Chen, B.; Dai, J.; Fang, Y.; Bai, Y.; Lin, Y.; Wei, H.; Zeng, X. C.; Huang, J., Defect passivation in hybrid perovskite solar cells using quaternary ammonium halide anions and cations. *Nat. Energy* **2017,** *2* (7), 1-9.

8. Zhao, Y.; Tan, H.; Yuan, H.; Yang, Z.; Fan, J. Z.; Kim, J.; Voznyy, O.; Gong, X.; Quan, L. N.; Tan, C. S., Perovskite seeding growth of formamidinium-lead-iodide-based perovskites for efficient and stable solar cells. *Nat. Commun.* **2018,** *9* (1), 1-10.

9. Alsalloum, A. Y.; Turedi, B.; Zheng, X.; Mitra, S.; Zhumekenov, A. A.; Lee, K. J.; Maity, P.; Gereige, I.; AlSaggaf, A.; Roqan, I. S., Low-temperature crystallization enables 21.9% efficient single-crystal MAPbI3 inverted perovskite solar cells. *ACS Energy Lett.* **2020,** *5* (2), 657-662.

10. Boyd, C. C.; Shallcross, R. C.; Moot, T.; Kerner, R.; Bertoluzzi, L.; Onno, A.; Kavadiya, S.; Chosy, C.; Wolf, E. J.; Werner, J., Overcoming redox reactions at perovskite-nickel oxide interfaces to boost voltages in perovskite solar cells. *Joule* **2020,** *4* (8), 1759-1775.

11. Hu, J.; Wang, C.; Qiu, S.; Zhao, Y.; Gu, E.; Zeng, L.; Yang, Y.; Li, C.; Liu, X.; Forberich, K., Spontaneously self-assembly of a 2D/3D heterostructure enhances the efficiency and stability in printed perovskite solar cells. *Adv. Energy Mater.* **2020,** *10* (17), 2000173.

12. Ansari, F.; Shirzadi, E.; Salavati-Niasari, M.; LaGrange, T.; Nonomura, K.; Yum, J.-H.; Sivula, K.; Zakeeruddin, S. M.; Nazeeruddin, M. K.; Grätzel, M., Passivation mechanism exploiting surface dipoles affords high-performance perovskite solar cells. *J. Am. Chem. Soc.* **2020,** *142* (26), 11428-11433.

13. Leguy, A. M.; Hu, Y.; Campoy-Quiles, M.; Alonso, M. I.; Weber, O. J.; Azarhoosh, P.; Van Schilfgaarde, M.; Weller, M. T.; Bein, T.; Nelson, J., Reversible hydration of CH3NH3PbI3 in films, single crystals, and solar cells. *Chem. Mater.* **2015,** *27* (9), 3397-3407.

14. Cho, Y.; Soufiani, A. M.; Yun, J. S.; Kim, J.; Lee, D. S.; Seidel, J.; Deng, X.; Green, M. A.; Huang, S.; Ho-Baillie, A. W., Mixed 3D–2D passivation treatment for mixed-cation lead mixed-halide perovskite solar cells for higher efficiency and better stability. *Adv. Energy Mater.* **2018,** *8* (20), 1703392.

15. Chen, Y.; Yu, S.; Sun, Y.; Liang, Z., Phase engineering in quasi-2D Ruddlesden–Popper perovskites. *J. Phys. Chem. Lett.* **2018,** *9* (10), 2627-2631.

16. Lai, H.; Kan, B.; Liu, T.; Zheng, N.; Xie, Z.; Zhou, T.; Wan, X.; Zhang, X.; Liu, Y.; Chen, Y., Two-dimensional Ruddlesden–Popper perovskite with nanorod-like morphology for solar cells with efficiency exceeding 15%. *J. Am. Chem. Soc.* **2018,** *140* (37), 11639-11646.

17. Blancon, J.-C.; Tsai, H.; Nie, W.; Stoumpos, C. C.; Pedesseau, L.; Katan, C.; Kepenekian, M.; Soe, C. M. M.; Appavoo, K.; Sfeir, M. Y., Extremely efficient internal exciton dissociation through edge states in layered 2D perovskites. *Science* **2017,** *355* (6331), 1288-1292.

18. Chen, J.; Seo, J. Y.; Park, N. G., Simultaneous improvement of photovoltaic performance and stability by in situ formation of 2D perovskite at (FAPbI3) 0.88 (CsPbBr3) 0.12/CuSCN interface. *Adv. Energy Mater.* **2018,** *8* (12), 1702714.

19. Mao, L.; Ke, W.; Pedesseau, L.; Wu, Y.; Katan, C.; Even, J.; Wasielewski, M. R.; Stoumpos, C. C.; Kanatzidis, M. G., Hybrid Dion–Jacobson 2D lead iodide perovskites. *J. Am. Chem. Soc.* **2018,** *140* (10), 3775-3783.

20. Chen, P.; Bai, Y.; Wang, S.; Lyu, M.; Yun, J. H.; Wang, L., In situ growth of 2D perovskite capping layer for stable and efficient perovskite solar cells. *Adv. Funct. Mater.* **2018,** *28* (17), 1706923.

21. Wetzelaer, G. J. A.; Scheepers, M.; Sempere, A. M.; Momblona, C.; Ávila, J.; Bolink, H. J., Trap-assisted non-radiative recombination in organic–inorganic perovskite solar cells. *Adv. Mater.* **2015,** *27* (11), 1837-1841.

22. Wang, Q.; Dong, Q.; Li, T.; Gruverman, A.; Huang, J., Thin insulating tunneling contacts for efficient and water-resistant perovskite solar cells. *Adv. Mater.* **2016,** *28* (31), 6734-6739.

23. Buin, A.; Pietsch, P.; Xu, J.; Voznyy, O.; Ip, A. H.; Comin, R.; Sargent, E. H., Materials processing routes to trap-free halide perovskites. *Nano Lett.* **2014,** *14* (11), 6281-6286.

24. Xiong, S.; Hou, Z.; Zou, S.; Lu, X.; Yang, J.; Hao, T.; Zhou, Z.; Xu, J.; Zeng, Y.; Xiao, W., Direct observation on p-to n-type transformation of perovskite surface region during defect passivation driving high photovoltaic efficiency. *Joule* **2021,** *5* (2), 467-480.

25. Guo, P.; Ye, Q.; Liu, C.; Cao, F.; Yang, X.; Ye, L.; Zhao, W.; Wang, H.; Li, L.; Wang, H., Double barriers for moisture degradation: assembly of hydrolysable hydrophobic molecules for stable perovskite solar cells with high open-circuit voltage. *Adv. Funct. Mater.* **2020,** *30* (28), 2002639.

26. Li, H.; Shi, J.; Deng, J.; Chen, Z.; Li, Y.; Zhao, W.; Wu, J.; Wu, H.; Luo, Y.; Li, D., Intermolecular π–π conjugation self-assembly to stabilize surface passivation of highly efficient perovskite solar cells. *Adv. Mater.* **2020,** *32* (23), 1907396.

27. Cho, K. T.; Grancini, G.; Lee, Y.; Oveisi, E.; Ryu, J.; Almora, O.; Tschumi, M.; Schouwink, P. A.; Seo, G.; Heo, S., Selective growth of layered perovskites for stable and efficient photovoltaics. *Energy Environ. Sci.* **2018,** *11* (4), 952-959.

28. Wang, Z.; Lin, Q.; Chmiel, F. P.; Sakai, N.; Herz, L. M.; Snaith, H., Efficient ambient-air-stable solar cells with 2D–3D heterostructured butylammonium-caesium-formamidinium lead halide perovskites. *Nat. Energy* **2017,** *2* (9), 1-10.

29. Yoo, J. J.; Wieghold, S.; Sponseller, M. C.; Chua, M. R.; Bertram, S. N.; Hartono, N. T. P.; Tresback, J. S.; Hansen, E. C.; Correa-Baena, J.-P.; Bulović, V., An interface stabilized perovskite solar cell with high stabilized efficiency and low voltage loss. *Energy Environ. Sci.* **2019,** *12* (7), 2192-2199.

30. Cho, K. T.; Zhang, Y.; Orlandi, S.; Cavazzini, M.; Zimmermann, I.; Lesch, A.; Tabet, N.; Pozzi, G.; Grancini, G.; Nazeeruddin, M. K., Water-repellent low-dimensional fluorous perovskite as interfacial coating for 20% efficient solar cells. *Nano Lett.* **2018,** *18* (9), 5467-5474.

31. Yu, D.; Wei, Q.; Li, H.; Xie, J.; Jiang, X.; Pan, T.; Wang, H.; Pan, M.; Zhou, W.; Liu, W., Quasi-2D Bilayer Surface Passivation for High Efficiency Narrow Bandgap Perovskite Solar Cells. *Angew. Chem. Int. Ed.* **2022,** *20*, e202202346.

32. Ni, C.; Huang, Y.; Zeng, T.; Chen, D.; Chen, H.; Wei, M.; Johnston, A.; Proppe, A. H.; Ning, Z.; Sargent, E. H. J. A. C., Thiophene Cation Intercalation to Improve Band-Edge Integrity in Reduced-Dimensional Perovskites. *Angew. Chem.* **2020,** *132* (33), 14081-14087.

33. Qin, Y.; Zhong, H.; Intemann, J. J.; Leng, S.; Cui, M.; Qin, C.; Xiong, M.; Liu, F.; Jen, A. K. Y.; Yao, K. J. A. E. M., Coordination Engineering of Single-Crystal Precursor for Phase Control in Ruddlesden–Popper Perovskite Solar Cells. *Adv. Energy Mater.* **2020,** *10* (16), 1904050.

34. Fu, L.; Li, H.; Wang, L.; Yin, R.; Li, B.; Yin, L. J. E., Defect passivation strategies in perovskites for an enhanced photovoltaic performance. *Energy Environ. Sci.* **2020,** *13* (11), 4017-4056.

35. Parashar, M.; Singh, R.; Yoo, K.; Lee, J.-J. J. A. A. E. M., Formation of 1-D/3-D fused perovskite for efficient and moisture stable solar cells. *ACS Appl. Energy Mater.* **2021,** *4* (3), 2751-2760.

36. Sutanto, A. A.; Drigo, N.; Queloz, V. I.; Garcia-Benito, I.; Kirmani, A. R.; Richter, L. J.; Schouwink, P. A.; Cho, K. T.; Paek, S.; Nazeeruddin, M. K. J. J. o. M. C. A., Dynamical evolution of the 2D/3D interface: a hidden driver behind perovskite solar cell instability. *J. Mater. Chem. A* **2020,** *8* (5), 2343-2348.

37. Sutanto, A. A.; Szostak, R.; Drigo, N.; Queloz, V. I.; Marchezi, P.; Germino, J.; Tolentino, H. C.; Nazeeruddin, M. K.; Nogueira, A. F.; Grancini, G., In situ analysis reveals the role of 2D perovskite in preventing thermal-induced degradation in 2D/3D perovskite interfaces. *Nano Lett.* **2020,** *20* (5), 3992-3998.

38. Lin, J. T.; Hu, Y. K.; Hou, C. H.; Liao, C. C.; Chuang, W. T.; Chiu, C. W.; Tsai, M. K.; Shyue, J. J.; Chou, P. T., Superior Stability and Emission Quantum Yield (23%±3%) of Single-Layer 2D Tin Perovskite $TEA_2SnI_4$ via Thiocyanate Passivation. *Small* **2020,** *16* (19), 2000903.

39. Wang, Z.; Wang, F.; Zhao, B.; Qu, S.; Hayat, T.; Alsaedi, A.; Sui, L.; Yuan, K.; Zhang, J.; Wei, Z., Efficient two-dimensional tin halide perovskite light-emitting diodes via a spacer cation substitution strategy. *J. Phys. Chem. Lett.* **2020,** *11* (3), 1120-1127.

40. Li, R.; Yi, C.; Ge, R.; Zou, W.; Cheng, L.; Wang, N.; Wang, J.; Huang, W., Room-temperature electroluminescence from two-dimensional lead halide perovskites. *Appl. Phys. Lett.* **2016,** *109* (15), 151101.

41. Lanzetta, L.; Marin-Beloqui, J. M.; Sanchez-Molina, I.; Ding, D.; Haque, S. A., Two-dimensional organic tin halide perovskites with tunable visible emission and their use in light-emitting devices. *ACS Energy Lett.* **2017,** *2* (7), 1662-1668.

42. Chen, M. Y.; Lin, J. T.; Hsu, C. S.; Chang, C. K.; Chiu, C. W.; Chen, H. M.; Chou, P. T. J., Strongly Coupled Tin-Halide Perovskites to Modulate Light Emission: Tunable 550–640 nm Light Emission (FWHM 36–80 nm) with a Quantum Yield of up to 6.4%. *Adv. Mater.* **2018,** *30* (20), 1706592.

43. Jiang, Z.; Chen, X.; Lin, X.; Jia, X.; Wang, J.; Pan, L.; Huang, S.; Zhu, F.; Sun, Z., Amazing stable open-circuit voltage in perovskite solar cells using AgAl alloy electrode. *Sol. Energy Mater. Sol. Cells* **2016,** *146*, 35-43.

44. Li, X.; Bi, D.; Yi, C.; Décoppet, J.-D.; Luo, J.; Zakeeruddin, S. M.; Hagfeldt, A.; Grätzel, M., A vacuum flash–assisted solution process for high-efficiency large-area perovskite solar cells. *Science* **2016,** *353* (6294), 58-62.

45. Noel, N. K.; Abate, A.; Stranks, S. D.; Parrott, E. S.; Burlakov, V. M.; Goriely, A.; Snaith, H., Enhanced photoluminescence and solar cell performance via Lewis base passivation of organic–inorganic lead halide perovskites. *ACS Nano* **2014,** *8* (10), 9815-9821.

46. Zhang, T.; Long, M.; Qin, M.; Lu, X.; Chen, S.; Xie, F.; Gong, L.; Chen, J.; Chu, M.; Miao, Q., Stable and efficient 3D-2D perovskite-perovskite planar heterojunction solar cell without organic hole transport layer. *Joule* **2018,** *2* (12), 2706-2721.

47. Du, Y.; Zhu, D.; Cai, Q.; Yuan, S.; Shen, G.; Dong, P.; Mu, C.; Wang, Y.; Ai, X.-C., Spacer Engineering of Thiophene-Based Two-Dimensional/Three-Dimensional Hybrid Perovskites for Stable and Efficient Solar Cells. *J. Phys. Chem. C* **2022,** *126* (7), 3351-3358.

48. Liu, Y.; Duan, J.; Zhang, J.; Huang, S.; Ou-Yang, W.; Bao, Q.; Sun, Z.; Chen, X. J. A. A. M. I., High efficiency and stability of inverted perovskite solar cells using phenethyl ammonium iodide-modified interface of NiOX and perovskite layers. *ACS Appl. Mater. Interfaces* **2019,** *12* (1), 771-779.

49.     Singh, R.; Sandhu, S.; Yadav, H.; Lee, J.-J. J. A. a. m. i., Stable triple-cation (Cs+–MA+–FA+) perovskite powder formation under ambient conditions for hysteresis-free high-efficiency solar cells. *ACS Appl. Mater. Interfaces* **2019,** *11* (33), 29941-29949.

50.     Lu, D.; Lv, G.; Xu, Z.; Dong, Y.; Ji, X.; Liu, Y., Thiophene-based two-dimensional Dion–Jacobson perovskite solar cells with over 15% efficiency. *J. Am. Chem. Soc.* **2020,** *142* (25), 11114-11122.

51.     Zheng, X.; Deng, Y.; Chen, B.; Wei, H.; Xiao, X.; Fang, Y.; Lin, Y.; Yu, Z.; Liu, Y.; Wang, Q., Dual functions of crystallization control and defect passivation enabled by sulfonic zwitterions for stable and efficient perovskite solar cells. *Adv. Mater.* **2018,** *30* (52), 1803428.

52.     Han, Q.; Bae, S. H.; Sun, P.; Hsieh, Y. T.; Yang, Y.; Rim, Y. S.; Zhao, H.; Chen, Q.; Shi, W.; Li, G., Single crystal formamidinium lead iodide (FAPbI3): insight into the structural, optical, and electrical properties. *Adv. Mater.* **2016,** *28* (11), 2253-2258.

53.     Nagaraju, N.; Kushavah, D.; Kumar, S.; Ray, R.; Gambhir, D.; Ghosh, S.; Pal, S. K. J. P. C. C. P., Through structural isomerism: positional effect of alkyne functionality on molecular optical properties. *Phys. Chem. Chem. Phys.* **2022,** *24* (5), 3303-3311.

54.     Sourabh, S.; Whiteside, V.; Sellers, I.; Zhai, Y.; Wang, K.; Beard, M. C.; Yeddu, V.; Bamidele, M.; Kim, D. J. P. R. M., Hot carrier redistribution, electron-phonon interaction, and their role in carrier relaxation in thin film metal-halide perovskites. *Phys. Rev. Mater.* **2021,** *5* (9), 095402.

55.     Hong, X.; Ishihara, T.; Nurmikko, A. V., Dielectric confinement effect on excitons in ${\mathrm{PbI}}_{4}$-based layered semiconductors. *Phys. Rev. B* **1992,** *45* (12), 6961-6964.

56. Lin, Q.; Armin, A.; Nagiri, R. C. R.; Burn, P. L.; Meredith, P., Electro-optics of perovskite solar cells. *Nat. Photonics* **2015,** *9* (2), 106-112.

57. Zhai, H.; Liao, F.; Song, Z.; Ou, B.; Li, D.; Xie, D.; Sun, H.; Xu, L.; Cui, C.; Zhao, Y., 2D PEA2PbI4–3D MAPbI3 Composite Perovskite Interfacial Layer for Highly Efficient and Stable Mixed-Ion Perovskite Solar Cells. *ACS Appl. Energy Mater.* **2021,** *4* (12), 13482-13491.

# Supporting Information

# Improvement of both performance and stability of photovoltaic devices by in situ formation of a sulfur-based 2D perovskite

*Milon Kundar [a,b], Sahil Bhandari [a,b], Sein Chung [c], Kilwon Cho [c], Satinder K. Sharma [d], Ranbir Singh[d*], and Suman Kalyan Pal [a,b*]*

[a]*School of Physical Sciences, India Institute of Technology Mandi, Kamand, Mandi-175005, Himachal Pradesh, India*

[b]*Advanced Materials Research Centre, India Institute of Technology Mandi, Kamand, Mandi-175005, Himachal Pradesh, India*

[c]*Department of Chemical Engineering, Pohang University of Science and Technology, Pohang 37673, South Korea*

[d]School of Computing and Electrical Engineering (SCEE), Indian Institute of Technology Mandi, Kamand, Mandi-175005, Himachal Pradesh, India

AUTHOR INFORMATION

**Corresponding Author**

*E-mail: ranbir.iitk@gmail.com, suman@iitmandi.ac.in; Phone: +91 1905 267040

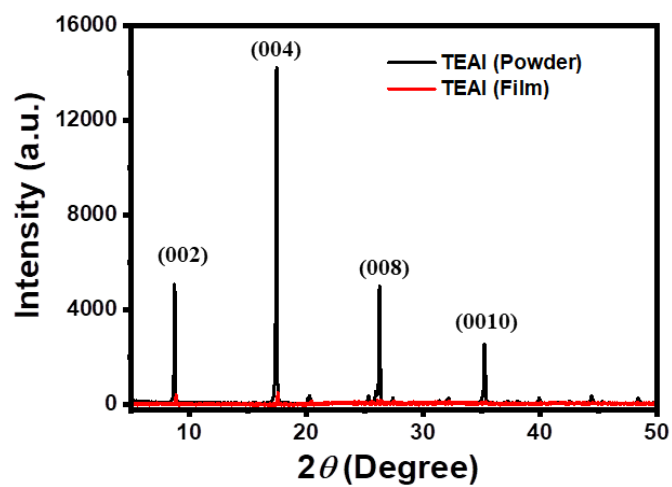

**Figure S1.** XRD patterns of the synthesized TEAI powder and film on glass substrates.

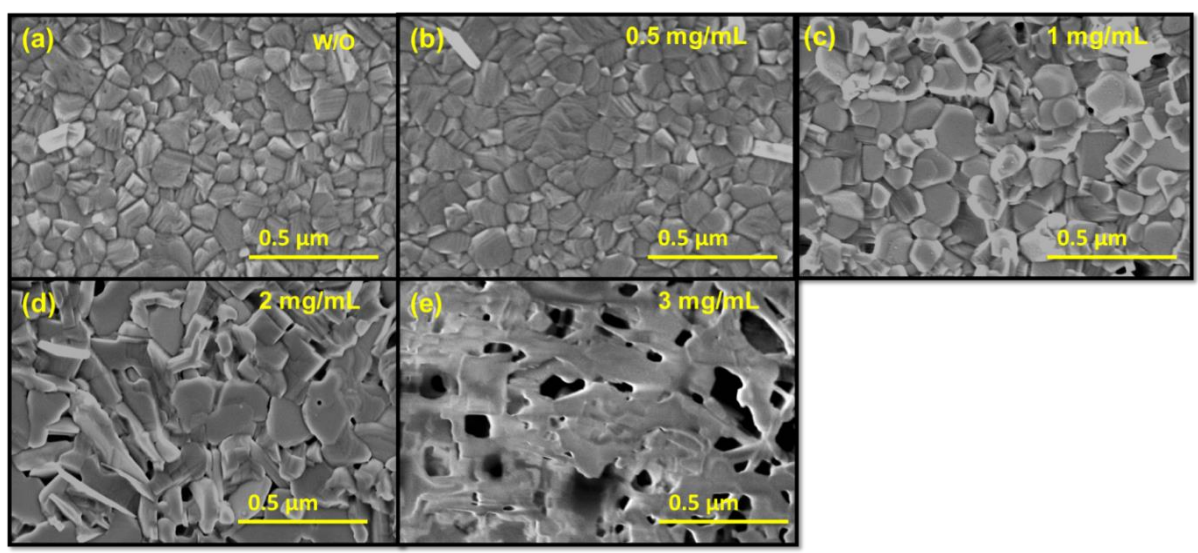

**Figure S2.** Top view of FESEM images of MAFA (3D) perovskite with different concentrations (0.5, 1, 2, and 3 mg mL$^{-1}$) of TEAI.

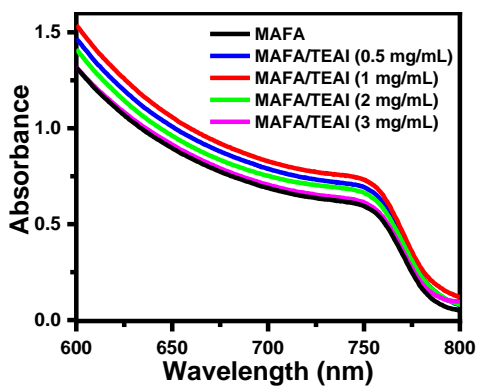

**Figure S3.** Absorption spectra of MAFA (3D) perovskite films with different concentrations (0.5, 1, 2, and 3 mg mL$^{-1}$) of TEAI.

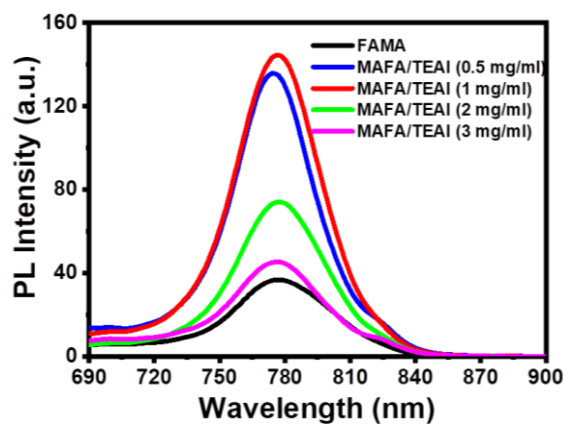

**Figure S4.** Photoluminescence (PL) spectra of MAFA (3D) perovskite films with different concentrations (0.5, 1, 2, and 3 mg mL$^{-1}$) of TEAI.

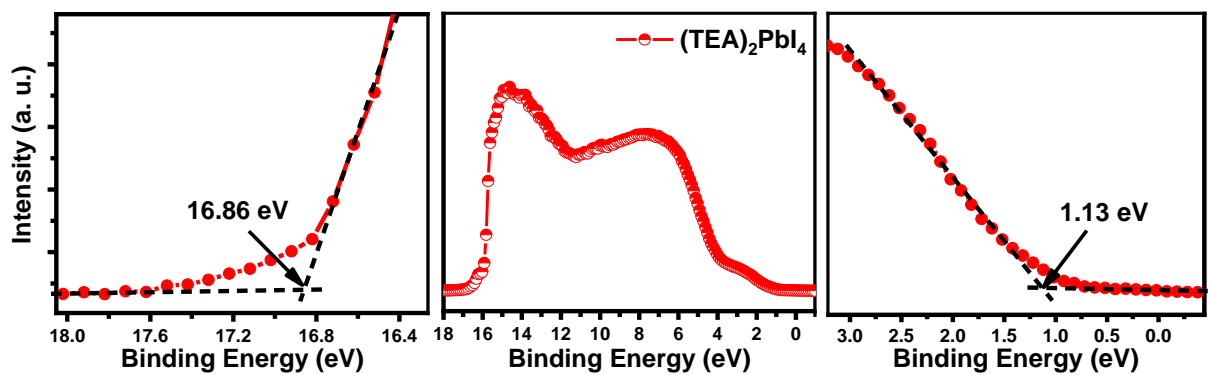

**Figure S5.** Ultraviolet photoelectron spectroscopy (UPS) spectra of 2D TEA$_2$PbI$_4$ perovskite film.

**Determination of energy level position of TEA$_2$PbI$_4$ perovskite**

Ultraviolet photoelectron spectroscopy (UPS) measurements were carried out to calculate the energy of the energy levels of the TEA$_2$PbI$_4$ perovskite. The work function (WF) was estimated by using the following equation[1]

$$\Phi = h\nu \,(21.22 \text{ eV}) - E_{cutoff} \qquad (1)$$

The secondary electron cut-off (E$_{cutoff}$) value is obtained from UPS measurement and found to be 16.86 eV (figure S5). The calculated value of WF for 2D TEA$_2$PbI$_4$ perovskite is 4.35 eV. It is clear from the UPS measurement (**Figure S5**) that VBM is located at 1.13 eV below the Fermi level (E$_F$). By subtracting the optical band gap (1.93 eV) from VBM (5.48 eV), we measured the CBM energy of 3.55 eV.

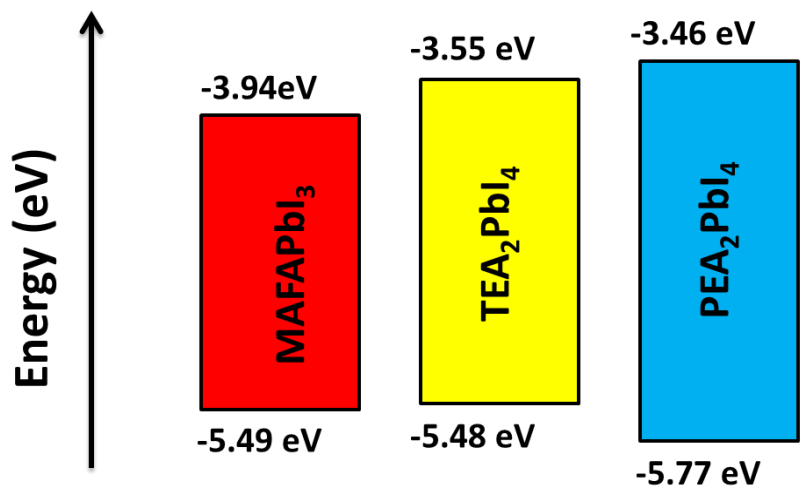

**Figure S6.** Energy level diagram of (TEA)$_2$PbI$_4$ and (PEA)$_2$PbI$_4$ perovskites. The energy of the levels of (PEA)$_2$PbI$_4$ is obtained from the literature.[2]

**Table S1**. Photovoltaic parameters of the PSCs fabricated with varied concentrations of TEAI

| Devices | V$_{OC}$ (V) | J$_{SC}$ (mA/cm$^2$) | FF (%) | PCE (%) |
|---|---|---|---|---|
| **MAFA** | 1.04 | 23.74 | 70.56 | 17.42 |
| **MAFA/TEAI (0.5mg/mL)** | 1.05 | 24.01 | 71.32 | 17.98 |
| **MAFA/TEAI (1mg/mL)** | 1.10 | 24.38 | 74.82 | 20.06 |
| **MAFA/TEAI (2mg/mL)** | 1.10 | 23.94 | 72.11 | 18.98 |
| **MAFA/TEAI (3mg/mL)** | 1.06 | 22.41 | 69.35 | 16.47 |

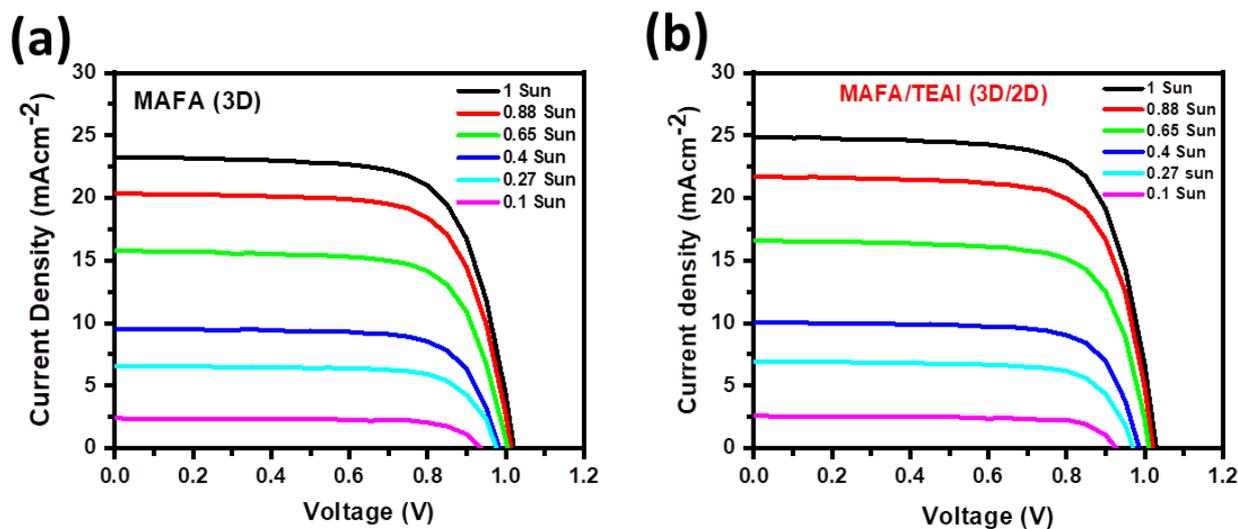

**Figure S7.** *J-V* responses of the fabricated PSCs (a) without and (b) with TEAI under light intensities from 100 mWcm$^{-2}$ (1 Sun) to 10 mWcm$^{-2}$ (0.1 Sun).

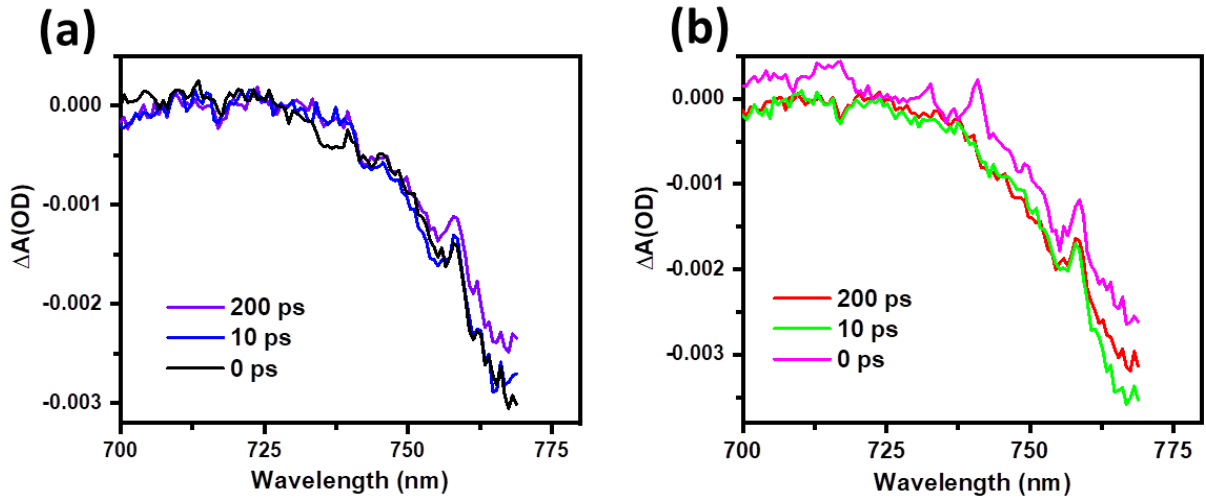

**Figure S8.** TA spectra of (a) MAFA (3D) and (b) MAFA/TEAI (3D/2D) perovskite films. Excitation wavelength is 532 nm.

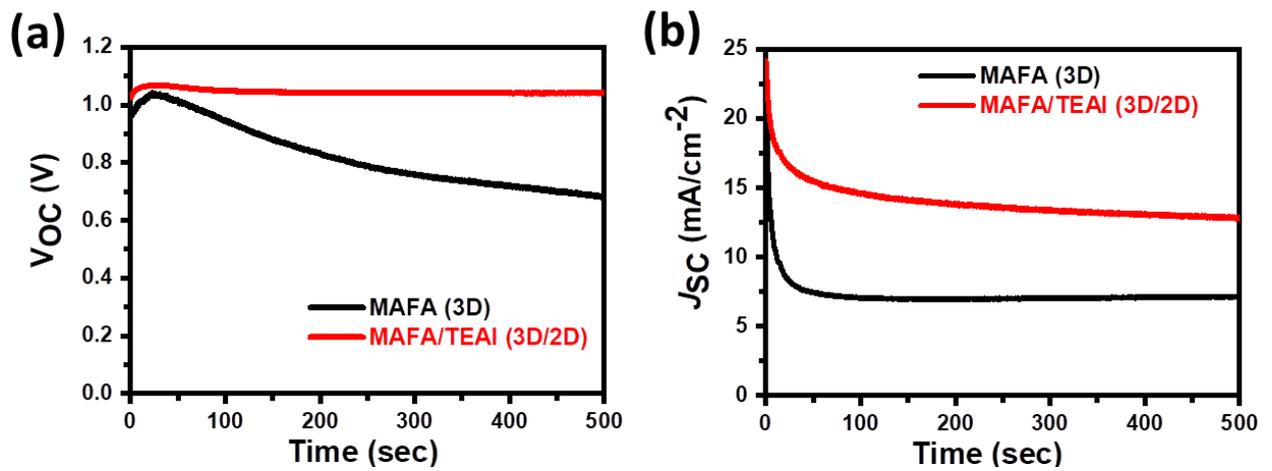

**Figure S9.** Time dependence of (a) open-circuit voltage ($V_{OC}$), and (b) short circuit current density ($J_{SC}$) under continuous white light illumination (100 mWcm$^{-2}$) in ambient condition (relative humidity (RH) ≈ 56% ± 4%).

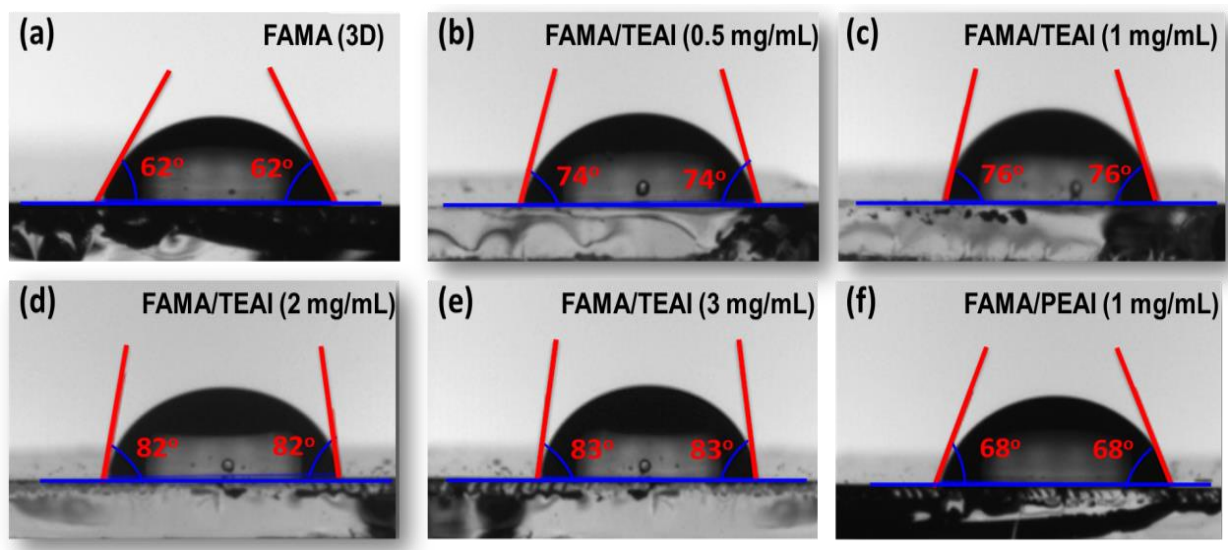

**Figure S10.** Images showing contact angles of water droplet on the surfaces of (a) MAFA (3D), (b) MAFA/TEAI (0.5 mg/mL), (c) MAFA/TEAI (1 mg/mL), (d) MAFA/TEAI (2 mg/mL), (e) MAFA/TEAI (3 mg/mL) and (f) MAFA/PEAI (1 mg/mL) perovskite films.

**References**

1.  Zheng, J.; Hu, L.; Yun, J. S.; Zhang, M.; Lau, C. F. J.; Bing, J.; Deng, X.; Ma, Q.; Cho, Y.; Fu, W. J. A. A. E. M., Solution-processed, silver-doped NiO x as hole transporting layer for high-efficiency inverted perovskite solar cells. *ACS Appl. Energy Mater.* **2018**, *1* (2), 561-570.

2.  Liu, Y.; Duan, J.; Zhang, J.; Huang, S.; Ou-Yang, W.; Bao, Q.; Sun, Z.; Chen, X. J. A. A. M. I., High efficiency and stability of inverted perovskite solar cells using phenethyl ammonium iodide-modified interface of NiOX and perovskite layers. *ACS Appl. Mater. Interfaces* **2019**, *12* (1), 771-779.